\documentclass{PoS}
\usepackage{amssymb, wrapfig} 

\title{Accretion Disk Reverberation Mapping of Active Galactic Nuclei at Wise Observatory}

\ShortTitle{Accretion Disk Reverberation Mapping of AGNs}

\author{Francisco Pozo Nu\~nez\\

Haifa Research Center for Theoretical Physics and Astrophysics, University of Haifa, 199 Aba Khoushy Ave. Mount Carmel, Haifa 3498838, Israel\\ 
E-mail:\email{francisco.pozon@gmail.com}}

\author{Doron Chelouche\\

Haifa Research Center for Theoretical Physics and Astrophysics, University of Haifa, 199 Aba Khoushy Ave. Mount Carmel, Haifa 3498838, Israel}

\author{Shai Kaspi\\

School of Physics and Astronomy and Wise Observatory, The Raymond and Beverly Sackler Faculty of Exact Sciences, Tel Aviv University, Tel Aviv 6997801, Israel}

\abstract{We have started automatized photometric monitoring of active galactic nuclei using the 46\,cm telescope of the Wise Observatory in Israel. The telescope is specially equipped with narrow-band filters to perform high-fidelity photometric reverberation mapping of the accretion disk in \textit{V} $<17$ mag sources up to $z \sim 0.1$. Here, we describe the capability and accuracy of the experiment, and present the first science verification data obtained for the Seyfert 1 galaxy Mrk\,279.

With sub-diurnal sampling over more than two months, and typical flux measurement uncertainties of $1\%$, we are able to measure inter-band time-delays of up to $\sim 2$ days across the optical range.}

\FullConference{Revisiting narrow-line Seyfert 1 galaxies and their place in the Universe - NLS1-2018\\
		9-13 April 2018 \\
		Padova Botanical Garden, Italy}

\begin{document}

\section{Introduction}

\begin{wrapfigure}[20]{r}{.45\textwidth}
  \centering
   \includegraphics[width=0.45\textwidth]{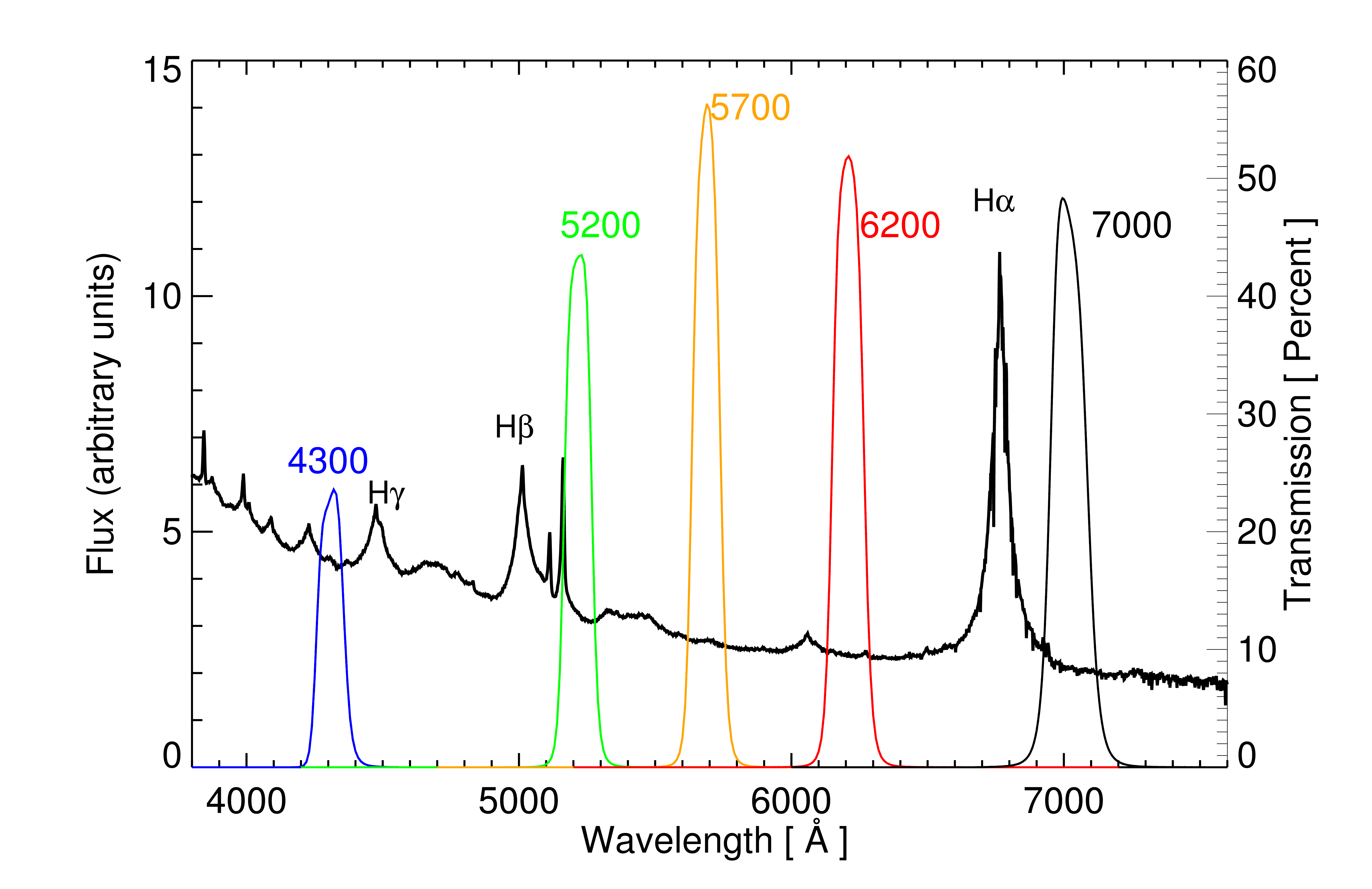}
  \caption{Effective transmission of the narrow-band filters. The AGN spectrum \cite{2006ApJ...640..579G} is shown at $z = 0.03$ where the emission line free continuum bands 4300, 5700, 6200 and 7000\,\AA\ trace the AGN continuum variations, suitable for photometric reverberation mapping of the accretion disk.\label{fig:filters}}
\label{filters}
\end{wrapfigure}

Reverberation mapping (RM) is the only method available to study the central engine of active galactic nuclei (AGNs) independent of the spatial resolution of the instrument. RM uses spectroscopic \cite{1993PASP..105..247P,
2012ApJ...755...60G} and photometric \cite{2012A&A...545A..84P,2012ApJ...747...62C} monitoring to measure the light travel time between the accretion disk (AD) and the broad line region (BLR) and/or hot dust. The hot AD produces a 
variable continuum emission, and this variability is observed with a time delay ($\tau_{BLR}=R_{BLR}/c$) in the studied broad emission lines of the BLR. Similarly, if the dust torus is located at some radial distance $R_{dust}$ 
around the hot AD it will reprocess the UV/optical radiation to thermal near-infrared (NIR) radiation with a characteristic time delay $\tau_{dust}=R_{dust}/c$.

A further use of spectroscopic and photometric RM is to map the AD. Depending on the local temperature of the AD, the time delays between light curves at different continuum bands can be interpreted as the light travel time ($\tau_{AD}$) across the AD \cite{1998ApJ...500..162C}. Therefore, as a first approximation, the time delays yield valuable information about the size ($R_{AD} \sim c \cdot \tau_{AD}$) and the temperature stratification across the AD, both crucial parameters provide important constraints on actual theoretical models of the AD.

Motivated by the recent progress in photometric RM we have started an automatized photometric monitoring of selected AGNs with brightness \textit{V} $<17^{m}$ up to $z \sim 0.1$. Here we present the first science verification data obtained for the Seyfert 1 galaxy Mrk\,279.

\section{Observations}
The photometric monitoring campaign started on March 2016 using the 46\,cm telescope of the Wise Observatory in Israel\footnote{https://physics.tau.ac.il/astrophysics/wiseobservatory}. Observations are currently performed in a fully robotic mode using several scripts integrated with the Astronomers Control Program (ACP\footnote{http://acp.dc3.com/}). For each night, the data are automatically transferred to the Hive computer cluster at the University of Haifa where they are reduced using IRAF\footnote{IRAF is distributed by the National Optical Astronomy Observatory, which is operated by the Association of Universities for Research in Astronomy (AURA) under cooperative agreement with the National Science Foundation.} packages and custom written tools. A detailed description of the telescope operation and data reduction process is presented in Pozo Nu\~nez et al. \cite{2017PASP..129i4101P}.

\begin{wrapfigure}[20]{r}{.45\textwidth}
  \centering
   \includegraphics[width=0.45\textwidth]{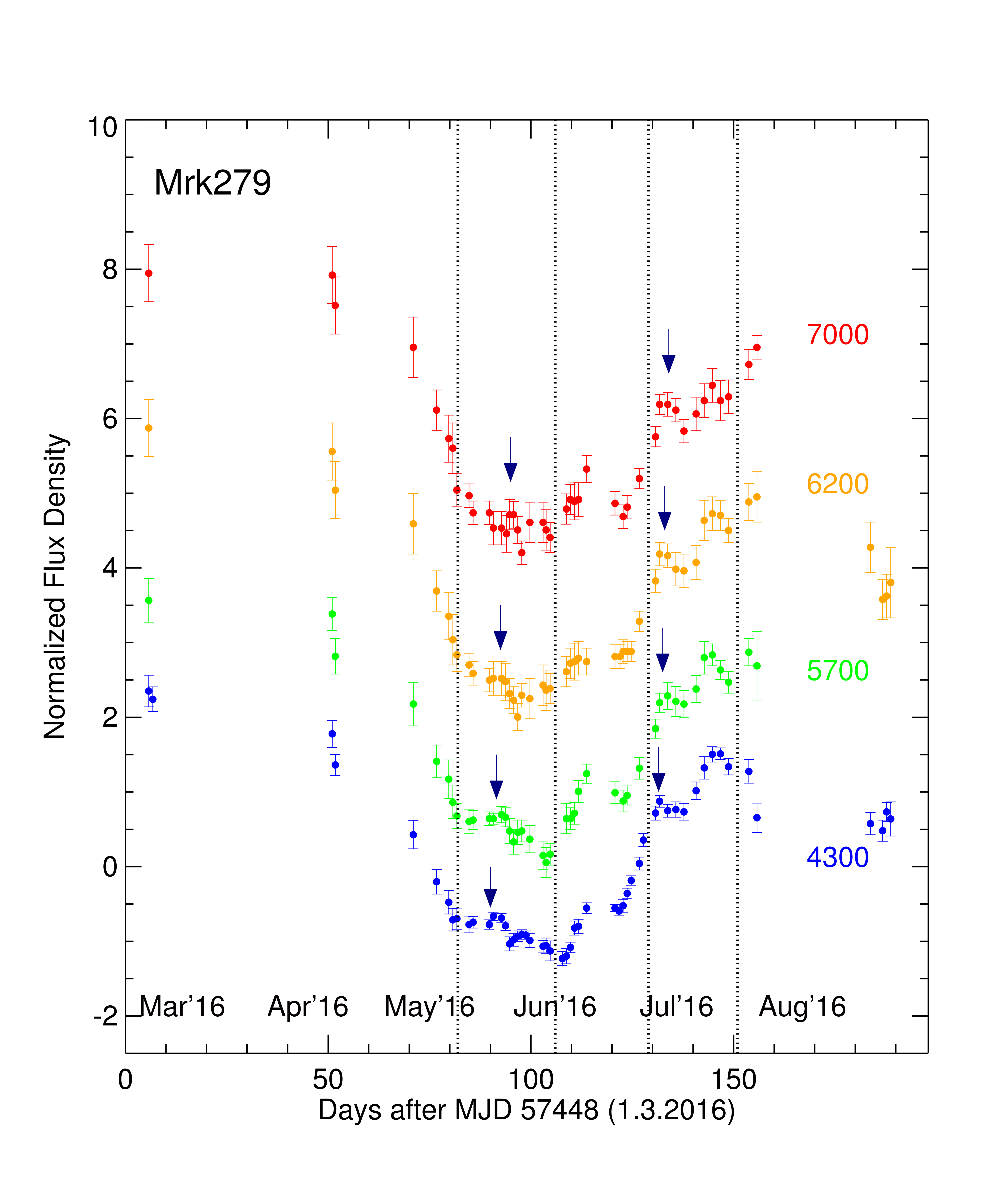}
  \caption{Observed light curves of the Seyfert 1 galaxy Mrk\,279. The dotted lines mark two windows of short variability events. The arrows mark the variation features for which a time delay of $\sim1.5$ days between the bands can be detected.}
\label{lc}
\end{wrapfigure}

Light curves are obtained with a median sampling of 0.5 to 1 days in the narrow bands (NB) 4300, 5200, 5700, 6200 and 7000\,\AA . The NB are carefully chosen in order to study only the emission line free AGN continuum (Figure 1). Photometry is performed using image subtraction techniques \cite{2000A&AS..144..363A}. For some AGNs, the number of stars in the field is not enough to achieve a reliable point-spread function kernel determination, therefore we extract the light curves using aperture photometry. The optimal aperture photometry is chosen to minimize the host galaxy contribution\footnote{More information about the photometric methods can be found in Pozo Nu\~nez et al. \cite{2017PASP..129i4101P}.}. Mrk\,279 lies at redshift $z = 0.0306$ and the bands 4300, 5700, 6200, and 7000\,\AA\ trace mainly the AGN continuum variations (Figure 1).

\section{Results}

The light curves of Mrk\,279 are shown in Figure 2. Observations were performed between March 2016 and August 2016. The long-term variations are well correlated among different bands and the quality of the light curves allows us to detect significant short timescale variations features. The variation observed during the variability windows  MJD57448+82--MJD57448+108, and MJD57448+129--MJD57448+151, allows us to detect a time delay between the continuum bands of about 2 days. Both 4300-band steep declines starting at MJD57448+90 and MJD57448+131 are roughly consistent with the 5700-band shifted by about 1.5 days. The same holds for the 6200 and 7000-bands during both variability windows. A more detailed study of the light curves of Mrk\,279, including cross-correlation analysis will be presented in a forthcoming publication.

\section{Summary}

\begin{figure*}
  \centering
  \includegraphics[width=9cm,height=6cm]{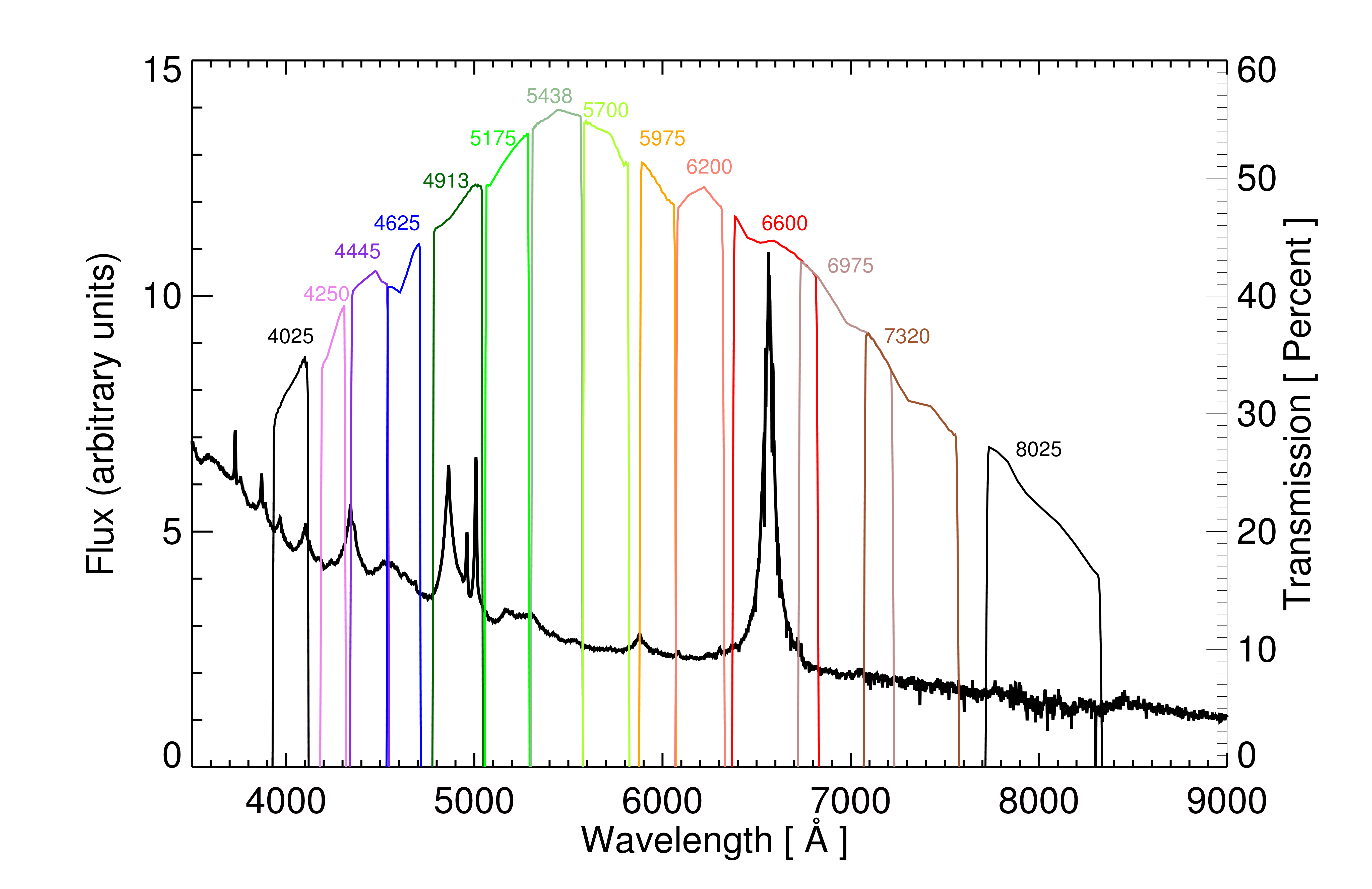}
  \caption{Same as Fig. \ref{fig:filters}, but for the newly commissioned 14 filters overlaid with the $z = 0$ composite AGN spectrum of Glikman et al. \cite{2006ApJ...640..579G}.\label{fig:nextfilt}}
\end{figure*}

We have presented the first science verification data of a new photometric RM monitoring of AGNs, using the robotic 46\,cm telescope of the Wise Observatory in Israel. The use of image subtraction and aperture photometry techniques allows us to measure the nuclear flux of the Seyfert 1 galaxy Mrk\,279 with high accuracy ($< 1\%$). We are able to measure inter-band time-delays of up to $\sim 2$ days across the optical range. 

We are commissioning a second CCD camera (QSI KAF-8300) with 14 new filters. The new filters will provide a better wavelength coverage of both line-free continuum regions and BLR H$\alpha$/H$\beta$ emission lines (Figure 3), hence decreasing the biases in the time-lag measurements of the AD and the BLR for several AGNs observed in the future.

\section*{Acknowledgements}

This conference has been organized with the support of the
Department of Physics and Astronomy ``Galileo Galilei'', the 
University of Padova, the National Institute of Astrophysics 
INAF, the Padova Planetarium, and the RadioNet consortium. 
RadioNet has received funding from the European Union's
Horizon 2020 research and innovation programme under 
grant agreement No~730562. This research has been partly supported by grants 950/15 from the Israeli Science Foundation (ISF) and 3555/14-1 from the Deutsche Forschungsgemeinschaft (DFG). Hive computer cluster at the university of Haifa is partly supported by ISF grant 2155/15.

\bibliographystyle{JHEP}
\bibliography{pozonunez}

\begin{thebibliography}{10}

\bibitem{2006ApJ...640..579G}
E.~{Glikman}, D.~J. {Helfand} and R.~L. {White}, \emph{{A Near-Infrared
  Spectral Template for Quasars}},
  \href{https://doi.org/10.1086/500098}{\emph{\apj} {\bfseries 640} (2006) 579}
  [\href{https://arxiv.org/abs/astro-ph/0511640}{{\ttfamily
  astro-ph/0511640}}].

\bibitem{1993PASP..105..247P}
B.~M. {Peterson}, \emph{{Reverberation Mapping of Active Galactic Nuclei}},
  \href{https://doi.org/10.1086/133140}{\emph{\pasp} {\bfseries 105} (1993)
  247}.

\bibitem{2012ApJ...755...60G}
C.~J. {Grier} et~al., \emph{{Reverberation Mapping Results for Five Seyfert 1
  Galaxies}}, \href{https://doi.org/10.1088/0004-637X/755/1/60}{\emph{\apj}
  {\bfseries 755} (2012) 60} [\href{https://arxiv.org/abs/1206.6523}{{\ttfamily
  1206.6523}}].

\bibitem{2012A&A...545A..84P}
F.~{Pozo Nu{\~n}ez} et~al., \emph{{Photometric Reverberation Mapping of 3C 120}},
  \href{https://doi.org/10.1051/0004-6361/201219107}{\emph{\aap} {\bfseries
  545} (2012) A84} [\href{https://arxiv.org/abs/1303.3506}{{\ttfamily
  1303.3506}}].

\bibitem{2012ApJ...747...62C}
D.~{Chelouche} and E.~{Daniel}, \emph{{Photometric Reverberation Mapping of the
  Broad Emission Line Region in Quasars}},
  \href{https://doi.org/10.1088/0004-637X/747/1/62}{\emph{\apj} {\bfseries 747}
  (2012) 62} [\href{https://arxiv.org/abs/1105.5312}{{\ttfamily 1105.5312}}].

\bibitem{1998ApJ...500..162C}
S.~J. {Collier} et~al., \emph{{Steps toward Determination of the Size and
  Structure of the Broad-Line Region in Active Galactic Nuclei. XIV. Intensive
  Optical Spectrophotometric Observations of NGC 7469}},
  \href{https://doi.org/10.1086/305720}{\emph{\apj} {\bfseries 500} (1998)
  162}.

\bibitem{2017PASP..129i4101P}
F.~{Pozo Nu{\~n}ez} et~al.,
  \emph{{Automatized Photometric Monitoring of Active Galactic Nuclei with the
  46cm Telescope of the Wise Observatory}},
  \href{https://doi.org/10.1088/1538-3873/aa7a55}{\emph{\pasp} {\bfseries 129}
  (2017) 094101} [\href{https://arxiv.org/abs/1706.05463}{{\ttfamily
  1706.05463}}].

\bibitem{2000A&AS..144..363A}
C.~{Alard}, \emph{{Image Subtraction Using a Space-Varying Kernel}},
  \href{https://doi.org/10.1051/aas:2000214}{\emph{\aaps} {\bfseries 144}
  (2000) 363}.

\end{thebibliography}

\end{document}